\documentstyle[12pt]{article}
\vfil
\begin{document}

\newpage
\pagestyle{empty}

\begin{flushright}

OUTP-96-53P 

CERN-TH/96-246 

hep-ph/9610202

September 1996

\end{flushright}

\vskip 0.6cm

\centerline{\large\bf Disoriented Chiral Condensates 
in Hadron-Hadron Collisions\footnote{Presented by G. Amelino-Camelia
at the {\it 10th International Conference on 
Problems of Quantum Field Theory},
Alushta, Crimea, Ukraine,  May 13-18, 1996.}}
\vskip 2 cm
\begin{center}
{\bf G. Amelino-Camelia$^{(a,b)}$, 
J.D. Bjorken$^{(a,c)}$,
and S.E. Larsson$^{(a)}$} \\
\end{center}
\begin{center}
{\it (a)  Theoretical Physics, 
University of Oxford, 1 Keble Rd., Oxford OX1 3NP, UK} \\
{\it (b) CERN, TH-Division, CH-1211 Geneva 23, Switzerland}\\
{\it (c) SLAC, Stanford University, Stanford, California 94309, USA}
\end{center}

\vskip 2cm
\centerline{\bf ABSTRACT }
We review recent progress in the description and understanding
of {\it disoriented chiral condensates}.
Certain important unsolved issues are underlined,
and the preliminary results of our program of investigation
of these issues in the framework of the
classical linear sigma model
are reported.
We also briefly review a formalism which could be useful
at the full non-equilibrium quantum field theory
level of analysis.

\vfill
\newpage
\baselineskip 12pt plus 0.2pt minus 0.2pt

\vskip 0.5 cm

\section{Introduction}
$~~~~\,$Recently,
in order to explain rare events with
a deficit or excess of neutral pions
observed in cosmic ray experiments,
there has been increased
interest in the conjecture that it
might be possible to
produce {\it disoriented chiral condensates} 
(DCCs), {\it i.e.}
correlated regions
wherein the quark condensate, $\langle 0| q_L  \bar{q}_R |0 \rangle$,
is chirally rotated from its usual orientation
in isospin space.

On the theoretical side there has been great interest
(see, {\it e.g.}, Refs. [1-14])
both in the development of technical tools suitable for the
description of this new phenomenon,
and in the exploration of the possibilities opened by DCCs
as probes of the structure of chromodynamics, 
most notably in relation to the chiral
phase transition.

The idea that such DCCs might be produced
in high energy collisions
at existing or planned hadron or heavy-ion accelerators
has originated
several experimental proposals.
In particular, one of us is co-spokesman for a Fermilab
experiment \cite{t864} 
looking for DCCs in hadron-hadron collisions.
In high energy $p$-${\bar p}$ collisions
which lead to a sizeable multiplicity of produced
particles, but not necessarily with high-$p_T$
jets in the final state,
the time evolution is quasi-macroscopic,
because the hadronization time can be rather large, 3-5$f$.
At times $t$ before hadronization,
the initial state partons, produced in a volume much smaller
than a cubic fermi,
stream outward at essentially the speed of light in
all directions, occupying the surface of a sphere
of radius $t$ (in units such that the speed of light is 1).
Most of the outgoing energy/momentum is expected to be concentrated
near the light cone, {\it i.e.} on the fireball surface.
However,
the interior of the fireball is also an interesting place.
If its energy density is low enough,
the interior should look very similar to the vacuum,
with an associated non-vanishing
quark condensate.
Since the energy density from the intrinsic chiral symmetry
breaking is small \cite{pisa,kota},
and the fireball surface isolates the interior
from the exterior of the light cone,
it is reasonable to consider the possibility that
well inside the light cone
the quark condensate might be
chirally rotated from its usual orientation.
At late times this disoriented vacuum
would relax back to ordinary vacuum,
radiating its collective modes, pions.
The properties of the radiated pions would be strongly affected by
the semiclassical, coherent nature of the process;
in particular, one could expect 
anomalously large
event-by-event fluctuations
in the ratio of the number of charged pions
to neutral pions produced.
Importantly, assuming that the 
event-by-event deviation of
the quark condensate from its usual orientation
be random,
one finds [1-4,16,17]
that the distribution $P(f)$ of the neutral 
fraction 
\begin{eqnarray}
f \equiv 
{N_{\pi^0} \over N_{\pi^0} + N_{\pi^+} + N_{\pi^-} }
\equiv 
{N_{\pi^0} \over N_{tot} }
~,\label{fdef}
\end{eqnarray}
is given by
\begin{eqnarray}
{dP \over df} = {1 \over 2 \, \sqrt{f}}
~,\label{radice}
\end{eqnarray}
at large $N_{tot}$.
Most notably, this implies that for ``DCC pions'' the probability of finding
extreme values of $f$ is very different from
ordinary pion production (which is given by a binomial distribution),
in which the fluctuations are expected to be peaked at $f \! = \! 1/3$
and fall exponentially away from the peak.
Recent experimental DCC searches \cite{t864} are largely based on
the structure of Eq. (\ref{radice}).

\section{Description at the Classical Level}
$~~~~\,$Since the DCC is essentially a classical pion field,
the description of its space-time evolution is most naturally described
using classical or semiclassical techniques.
The $O(4)$ $\sigma$-model
is typically used
as a model of chromodynamics in DCC studies.
The $\sigma$-model is simple enough to be treatable,
has the correct
chiral symmetry properties,
and describes the low energy phenomenology of pions.
The Lagrangian of the (linear)
$\sigma$-model is
(in the chiral limit $m_\pi \! = \! 0$)
\begin{eqnarray}
{\cal L} = {1 \over 2} (\partial_\mu \sigma)^2
+ {1 \over 2} (\partial_\mu \vec{\pi})^2
- {\lambda \over 4} (\sigma^2+\vec{\pi}^2 - f_\pi^2)^2
~.\label{lagrlin}
\end{eqnarray}

A meaningful $\sigma$-model description can start at some small proper
time, of order 0.2-0.3$f$, near the light cone,
when the collective coordinates
$\sigma$ and $\pi$ become relevant \cite{bjminn,gm}.
At this early proper time the distribution of the chiral field
\begin{eqnarray}
\Phi \equiv  \sigma +  i \vec{\pi} \vec{\tau}
~,\label{udef}
\end{eqnarray}
can be expected to be noisy, but with $\langle \! \Phi \! \rangle = \! 0$.

As proper time increases the field $\Phi$ rolls into
a minimum with $\Phi^+ \Phi \! = \! f_\pi^2$,
and during this ``rolling phase'' 
the pion mass is imaginary, leading to unstable growth
of the Goldstone modes \cite{krishna,bjminn}.
Since, as mentioned in the Introduction,
the energy density from the intrinsic chiral symmetry
breaking is small \cite{pisa,kota},
and the fireball surface isolates the interior
from the exterior of the light cone,
it is reasonable to expect that
the interior ends up in a disoriented vacuum.

At late times one such region of disoriented vacuum with
a given isospin orientation
would relax back to ordinary vacuum $\langle \! \Phi \! \rangle 
= \langle \! \sigma \! \rangle = \! f_\pi$,
radiating pions with the same isospin orientation.

In modelling these stages of evolution, the chiral limit can be
safely taken as long as the proper time
is small compared to $m_\pi^{-1}$,
while at times of order 1-2$m_\pi^{-1}$
the pion mass can no longer be neglected
and
one should \cite{bjminn}
decompose the DCC field into physical-pion normal modes
and let them propagate out to infinity as free states.

It is also reasonable to expect that
approximations
based on
the replacement of the full
linear $\sigma$-model (\ref{lagrlin})
by the simpler non-linear $\sigma$-model
\begin{eqnarray}
{\cal L} = {1 \over 2} (\partial_\mu \sigma)^2
+ {1 \over 2} (\partial_\mu \vec{\pi})^2
~,~~~ \rm{with} ~ \sigma^2 + \vec{\pi}^2 = f_\pi^2
~, \label{lagrnonlin}
\end{eqnarray}
could be reliably used at times late enough for
the chiral field to have
already rolled into a minimum with $\Phi^+ \Phi \! \equiv \!
\sigma^2 + \vec{\pi}^2 \! = \! f_\pi^2$.

Actually the entire picture of DCC evolution given above
can be implemented
(although, at least at early times,
the quantitative aspects are not accurately
evaluated)
within the non-linear $\sigma$-model
in the framework of the set of classical solutions
identified by Anselm \cite{anselm} and others \cite{blaizot,kota},
which have the form
\begin{eqnarray}
\Phi = f_\pi \, V_L^+ e^{ i \theta \tau_3} V_R
~,\label{urotaz}
\end{eqnarray}
where $V_L$ and $V_R$ are constant but otherwise
arbitrary matrices \cite{anselm},
and $\theta$ is such that 
\begin{eqnarray}
\Box \theta = 0 
~.\label{eqsteta} 
\end{eqnarray}

In particular, by taking $V_L \! = \! V_R$ and using
the following solution of (\ref{eqsteta})
\begin{eqnarray}
\theta(r) &=& {\pi \over 4} \, {2T+r-t \over 
r} \, \Theta(2T+r-t) \, \Theta(t-r)
\nonumber\\
& &- {\pi \over 4} \, {2T-r-t \over 
r} \, \Theta(2T-r-t) \, \Theta(t+r)
~,\label{exactlin}
\end{eqnarray}
where $\Theta$ is the step function,
one can describe several stages of the evolution of an ideal
spherically symmetric DCC formed at the collision centre.
For $t \leq 2T$ the DCC field described by (\ref{exactlin})
expands at the speed of light
(mimicking the ``rolling phase'' of the DCC evolution).
For $2T \leq t \leq 4T$ the DCC field keeps expanding
but the true vacuum starts breaking into the interior of the light cone.
For $t \geq 4T$
the true vacuum
is everywhere apart from 
a bump of DCC field propagating outward.
The final stage of evolution, the one in which the classical DCC
field radiates pions, cannot be reproduced within (\ref{exactlin});
it requires, as stated above, taking into account the non-vanishing
pion mass.

There are two more aspects of the simple solution (\ref{exactlin})
which are worth emphasizing. First we notice that (\ref{exactlin})  
is a solution of (\ref{eqsteta}) in the presence of a DCC source term
on the light cone at small times; such a source term is a straightforward
way to mimic
the unstable growth
of the Goldstone modes associated with the ``rolling phase'',
and could also be useful in more detailed analyses.
[Note that taking $V_L \! = \! V_R$ 
is very important, since otherwise
the source term
on the light cone will not vanish at large times.]
A second aspect of (\ref{exactlin}) which could be instructive
for other developments is the dominant role played by
the Minkowski geometry.
This remark also gives us a chance to state
our belief that geometry should play a central role
in DCC physics, and therefore the investigation of the possibility
of DCC formation in hadron-hadron collisions 
should proceed differently from the corresponding heavy-ion case.

Going back to the study of the space-time evolution of the DCC,
let us now consider the full
linear $\sigma$-model, which is necessary
for a description starting at early proper times.
A good starting point is the
generalization
to the
linear $\sigma$-model
of the
solutions (\ref{urotaz}) and (\ref{eqsteta})
of the non-linear $\sigma$-model.
One finds solutions of the form \cite{anselmnew}
\begin{eqnarray}
\Phi = \rho  \, V^+ e^{ i \theta \tau_3} V
~,\label{unew}
\end{eqnarray}
with $\theta$ and $\rho$ such that 
\begin{eqnarray}
\Box \theta = 0 ~,~~~
\Box \rho - \rho \, (\partial_\mu \theta)^2
+ \lambda \, \rho \, (\rho^2 - f_\pi^2) = 0 
~.\label{eqsnew}
\end{eqnarray}
The equation for $\rho$ is not easy to handle analytically,
but progress can be achieved by numerical techniques.

We are at present using numerical and analytic techniques
in a study of the space-time evolution of DCCs
within
the linear $\sigma$-model (\ref{lagrlin}).
The simulations start
from a given configuration for the pion/sigma fields
at an initial proper time,
and we observe the dynamics of the
transition to the ordered broken symmetry phase
as the system expands.
We plan to study different types of 
initial conditions, with the chiral field starting
either from the top  
or from the brim of the ``Mexican hat''
and with various assignments for the 
initial velocities in the
$\vec{\pi}$ and $\sigma$ directions.
One of the issues that we are interested in investigating
is the possibility of 
correlations in the evolution of two (or more) pieces of DCC
in different regions of the lego plot.
It would be interesting to check whether
these pieces of DCC ``attract'' or ``repel'',
and whether the presence of DCC in one region of the
lego plot stimulates production in neighbouring regions.
Ultimately, we hope
to create
simulations
realistic enough to be used
in the
interpretation of hadron-hadron experimental results relevant for
the DCC, but at present 
our research program is still in its preliminary stages.
We have started by investigating the linear $\sigma$-model
evolution of a
spherically symmetric DCC formed at the collision centre,
and observed that, as expected \cite{thn,bjminn}, the evolution
proceeds through the same stages of the simple non-linear $\sigma$-model
solution associated with (\ref{exactlin});
however, also depending on the initial conditions,
we find some evidence of a possible secondary wave front
resulting from implosion of the true vacuum as it occupies the
interior region of the fireball.
Such a secondary wave could in principle be very important
for experimental purposes, but much more numerical work is needed
to check whether in fact it can be expected to occur.
Other preliminary results concern simulations
of  scenarios with two pieces of DCC
and, for certain initial conditions,
they indicate that the evolution of one piece of DCC
is substantially affected by the presence of the other piece.
Also for this effect
much more numerical work is needed
in order to establish its
significance
for experimental purposes.
Our analytic and numerical work is also aimed
at describing,
within the proper framework of
Minkowskian geometry,
the way a piece of DCC affects
the evolution of another piece of DCC by acting as a source.

\section{Quantum Field Theory}
$~~~~\,$The expression in quantum terms 
of the ideas discussed classically in the previous section 
is most natural in the framework
of coherent states or squeezed states.
We refer the reader to Refs. \cite{cohe,ian,karma,kota} 
for analyses within these formalisms.
Here we
will briefly review
a formalism that could be useful in investigating
whether such quantum states are actually
created as a consequence of the
relevant
non-equilibrium quantum
dynamics.
We are not aware of much progress in the direction of exploiting
this formalism in the investigation of DCCs, at least in association
with hadron-hadron collisions\footnote{Related studies have been
reported in Refs. \cite{daniel,emilfred};
however, the formalism adopted there was somewhat different.
In particular,
the Liouville-vonNeumann equation (which in this section
will arise only after a suitable approximation)
was taken as a starting point.
Moreover, as a result of the assumptions made
about the geometry of the problem,
previous DCC analyses
using non-equilibrium quantum field theory are
more relevant
to the case of heavy-ion collisions.}, 
but a few conceptual points can be
made even before actual calculations.

One can describe a physical system
via its density matrix $\rho$, $\rho \equiv e^{- \beta H} /
(tr \, e^{- \beta H})$,
and average values of observables $O$ are
determined by the density matrix:
$\langle O \rangle \! = \!tr  \rho O $.
In equilibrium physics
the density matrix is time-independent;
in general, however, the density matrix is time-dependent,
and the task
of non-equilibrium quantum field theory is to
study
the time evolution of $\rho$.
Non-equilibrium quantum physics is a vast subject,
and there is no canonical approach to its investigation. 
Usually the approach is suggested by the specifics
of the physical system that one wants to describe.
We review an approach \cite{piebo,jackbanf}
that has proved
useful in early Universe cosmology \cite{guthpi}, and 
might also be useful in the investigation of DCCs.
This approach is set up in the
framework of the field theoretic Schr\"{o}dinger picture,
which is particularly suitable to time-dependent problems
that require an initial condition for a specific solution.
The (functional) density matrix is given by a superposition
of wave functionals 
\begin{eqnarray}
\rho(\phi_1,\phi_2) = \sum_n p_n \Psi_n(\phi_1)
\Psi^*_n(\phi_2) 
~,\label{eq3}
\end{eqnarray}
where $\{ \Psi_n \}$ is a complete set of wave functionals, 
and $p_n$ is the probability ($\sum_n p_n =1$)
that the system is in the state $\Psi_n$.
In general the $\Psi_n$'s and the $p_n$'s are time-dependent.

In equilibrium the dynamics is time-translation invariant and
energy is conserved.
Then the complete set of wave functionals $\{ \Psi_n \}$ can be chosen
to be the set of 
the (time-dependent) energy eigenstates,
while the $p_n$'s are time-independent and are given by the canonical
Boltzmann distribution, $p_n = e^{- \beta E_n} / (\sum_n e^{- \beta E_n})$,
where $E_n$ is the energy eigenvalue of the state $\Psi_n$.
The time evolution of the
corresponding density matrix is trivial: it remains constant 
in time because the $p_n$'s  and the $\Psi_n \Psi_n^*$'s
are constant (N.B.: the time dependence of the $\Psi_n$'s is just a phase).

For non-equilibrium physics the time evolution of the density matrix is 
instead non-trivial. In fact, it might not be possible to choose
the $\Psi_n$'s as energy eigenstates,
and the $p_n$'s need not be Boltzmann factors
and can change in time.
Under the assumption that the time dependence of the $\Psi_n$'s 
be determined by a time-dependent Schr\"{o}dinger equation, 
the density matrix $\rho$ satisfies the following
differential equation
\begin{eqnarray}
{d \rho \over dt} = \sum_n p_n {d \over dt} (\Psi_n \Psi_n^*)
+\sum_n {d p_n \over dt} (\Psi_n \Psi_n^*)
= i [ \rho , H] 
+\sum_n {d p_n \over dt} (\Psi_n \Psi_n^*)
~.\label{tero}
\end{eqnarray}
In order for Eq. (\ref{tero}) 
to describe a well defined initial value 
problem for the time evolution of $\rho$, 
it is necessary to give the form of $H$ and a model
for $dp_n/dt$.

One important simplification is allowed
when the evolution of interest is 
entropy-
conserving; in fact, in this case one can work with
time-independent $p_n$'s,
which indeed can be shown \cite{piebo}
to correspond to 
entropy-conserving time evolution.
In these hypotheses, Eq. (\ref{tero}) takes the form of the quantum
Liouville-vonNeumann equation with time-dependent Hamiltonian:
\begin{eqnarray}
{d \rho \over dt} = i [ \rho , H] 
~.\label{lvn}
\end{eqnarray}
While the investigation of the 
Liouville-von Neumann equation
is still extremely difficult, it is certainly remarkably
simpler than the original Eq. (\ref{tero}).
[We are not aware of any fruitful investigations of 
Eq. (\ref{tero}).]
It is therefore quite crucial to establish
whether the assumption of
entropy-conserving evolution
(time-independent $p_n$'s)
is justified in the study of DCCs.
In the {\it baked Alaska} scenario
for the formation of DCC in hadron-hadron collisions
that has been considered here, it appears
that the assumption
of entropy-conserving evolution 
could be quite accurate;
in fact, the interior of the fireball is a rather ``peaceful'' place.
Instead, in the case of heavy-ion collisions
the relevant time scales are such that the formation
of DCCs might be affected by non-entropy-conserving 
stages of the evolution of the quark-gluon plasma
({\it i.e.} one might be led to a definition
of {\it the system} such that the evolution
is entropy non-conserving).
Even when it is safe to assume that the time evolution is
entropy-conserving, and therefore
the Liouville-von Neumann equation can be 
meaningfully taken as a starting point,
for non-trivial Hamiltonians,
such as the ones relevant for DCC physics,
progress requires further approximations.
Interestingly, one can obtain approximate solutions
using the observation \cite{piebo,jackbanf}
that the Liouville-von Neumann equation can be
derived by varying an action-like
quantity within a variational principle introduced by Balian and 
Veneroni \cite{bave}.
An approximate application of the variational principle, with a restricted
variational {\it ansatz}, leads to approximate equations
for the density matrix\footnote{Alternatively,
as recently observed in Ref.\cite{emilfredverynew}, 
an {\it ansatz}
can be made directly at the level of the Liouville-von Neumann equation,
without advocating the Balian-Veneroni
variational riformulation of the problem.}.

\section{Outlook}
$~~~~\,$The recent increased interest in DCC physics has led
to substantial progress, but much work is still needed.
The challenge is quite strong both for
theorists, who must find their way through the complicated
models involved and provide experimentalists with a
reasonable (quantitative!) picture of what to look for,
and for experimentalists, who must disentangle the
(possibly faint) DCC signal from substantial
backgrounds.

We are confident that our program of numerical and analytical
investigation of the linear $\sigma$-model will soon
lead to
simulations 
realistic enough to be used
in the
interpretation of hadron-hadron experimental results relevant for
the DCC.

\vglue 0.6cm
\leftline{\Large {\bf Acknowledgements}}
\vglue 0.4cm
Much ot this work was initiated some time ago, before the onset of the
MiniMax experiment together with C. Taylor and K. Kowalski, which have
made similar investigations independently.
It is a pleasure to acknowledge conversations 
with D. Boyanovsky,
O. Eboli, F. Cooper, R. Jackiw, E. Mottola, S.-Y. Pi, 
K. Rajagopal, S. Sarkar, and P. White.
This work was
supported in part by funds provided by 
the European Union under contract \#ERBCHBGCT940685, 
the American Trust for Oxford University (George Eastman Visiting
Professorship), 
CSN and KV (Sweden), ORS and OOB (Oxford), and the Sir Richard
Stapley Educational Trust (Kent).

\baselineskip 12pt plus .5pt minus .5pt

\end{document}